\documentclass[aps,prl,twocolumn,showpacs,preprintnumbers,superscriptaddress]{revtex4}

\usepackage{graphicx}

\def\ltwid{\mathrel{\raise.3ex\hbox{$<$\kern-.75em\lower1ex\hbox{$\sim$}}}}
\def\gtwid{\mathrel{\raise.3ex\hbox{$>$\kern-.75em\lower1ex\hbox{$\sim$}}}}

\begin{document}

\preprint{UFIFT-QG-10-03}

\title{The $\zeta$-$\zeta$ Correlator Is Time Dependent}

\author{E. O. Kahya}
\email{emre-onur.kahya@uni-jena.de}
\affiliation{Theoretisch-Physikalisches Institut, 
             Friedrich-Schiller-Universit\"at Jena, Max-Wien-Platz 1, 
             D-07743 Jena, GERMANY}

\author{V. K. Onemli}
\email{onemli@itu.edu.tr}
\affiliation{Department of Physics, Istanbul Technical University,
             Maslak, Istanbul 34469, TURKEY}

\author{R. P. Woodard}
\email{woodard@phys.ufl.edu}
\affiliation{Department of Physics, University of Florida,
             Gainesville, FL 32611, USA}

\begin{abstract}

We comment on the recent arguments by Senatore and Zaldarriaga that
loop corrections to the $\zeta$-$\zeta$ correlator cannot grow with 
time after first horizon crossing. We first emphasize the need to
search for such secular dependence in corrections whose in-out matrix
elements are infrared singular on an infinite spatial manifold.
Then we give examples of such time dependence from pure quantum 
gravity and from scalar potential models. Finally, we point out that
this time dependence arises from inflationary particle production and
is therefore unlikely to endanger the preservation of super-horizon 
correlations as a record of inflation.

\end{abstract}


\maketitle

\section{Introduction}

Perhaps the most commonly quoted result for models of inflation is the
curvature power spectrum \cite{LL},
\begin{equation}
\Delta^2_{\mathcal{R}}(k,t) \equiv \frac{k^3}{2 \pi^2} \int \!\! d^3x \,
e^{-i \vec{k} \cdot \vec{x}} \Bigl\langle \Omega \Bigl\vert 
\mathcal{R}(t,\vec{x}) \mathcal{R}(t,\vec{0}) \Bigr\vert \Omega \Bigr\rangle 
\; .
\end{equation}
(The field $\mathcal{R}$ is defined by stripping the derivatives from
the 3-curvature in the co-moving frame for which the momentum flux vanishes
\cite{LL}.) These predictions are made in the context of perturbation 
theory about a homogeneous, isotropic and spatially flat geometry,
\begin{equation}
ds^2 = -dt^2 + a^2(t) d\vec{x} \cdot d\vec{x} \; \Rightarrow \;
H \equiv \frac{\dot{a}}{a} \; , \; \epsilon \equiv -\frac{\dot{H}}{H^2} .
\end{equation}
The time of first horizon crossing is $t_k$ such that $k = H(t_k) a(t_k)$,
after which $\Delta^2_{\mathcal{R}}(k,t)$ becomes nearly constant and one
drops the argument $t$. The tree order result for typical single-scalar 
inflation models is \cite{reviews},
\begin{equation}
\Delta^2_{\mathcal{R}}(k) \approx \frac{G H^2(t_k)}{\pi \epsilon(t_k)} \; .
\label{typR}
\end{equation}

Theorists are eager to predict $\Delta^2_{\mathcal{R}}(k)$ because its value 
for cosmological wave lengths can be reconstructed from observations of 
anisotropies in the cosmic microwave radiation and from large scale structure 
surveys \cite{WMAP},
\begin{equation}
\Delta^2_{\mathcal{R}}(k) = \Bigl(2.441^{+0.088}_{-0.092}\Bigr) 
\times 10^{-9} \Bigl( \frac{k}{.002~{\rm Mpc}^{-1}}\Bigr)^{-0.037 \pm 0.012}
\label{data}
\end{equation}
This connection between quantum gravity and cosmological observation
represents one of the great triumphs of inflation theory \cite{SMC}, and
accords expression (\ref{data}) the status of the first quantum gravitational 
data ever obtained.

Tree order results such as (\ref{typR}) derive from the linearized mode
functions. There can be important contributions from times before $t_k$ 
\cite{WMS}, but the mode functions become constant afterwards because the 
restoring force $k^2/a^2(t)$ redshifts away while the friction term remains 
large. Quantum loop effects can induce late time dependence by coupling 
mode $k$ to changes in the vacuum energy from the quantum fluctuations of 
other modes. A theorem by Weinberg limits this time dependence to powers of 
the ``infrared logarithm'', $\ln[a(t)/a(t_k)]$ \cite{SW1,SW2,Bua}. No one 
disputes this bound, the issue is its saturation.

In his first paper on the subject Weinberg considered two one loop 
processes which seemed to contribute infrared logarithms \cite{SW1}:
\begin{itemize}
\item{Section V gave a qualitative treatment of self-interactions within 
the gravity-inflaton system, culminating in equation (41); and}
\item{Section VII gave a computation of the contribution from $\mathcal{N}$
free, massless, minimally coupled scalars, culminating in equation (71).}
\end{itemize}
Although other work has produced similar results \cite{AEL,many}, Senatore 
and Zaldarriaga have argued that there cannot be any infrared logarithms 
from Weinberg's second source \cite{LSMZ}. We agree --- indeed, this follows 
from a simple rule for counting infrared logarithms \cite{TW1}. However, we 
do not accept the subsequent conclusion by Senatore and Zaldarriaga that 
$\Delta^2_{\mathcal{R}}(k,t)$ is free of infrared logarithms from any 
source and to all orders. (A recent paper by Giddings and Sloth also 
disputes their conclusion \cite{SGMS}.) The purpose of this paper is to 
show that infrared logarithms arise at one loop from self-interactions of 
the gravity-inflaton system (as in Weinberg's first example) and at two 
loops from massless, minimally coupled scalars with a quartic potential. 

In section 2 we summarize the Lagrangian. Section 3 describes a simple rule
for counting the maximum number of infrared logarithms which can derive from
a given interaction \cite{TW1}. In section 4 we compute a one loop effect 
from self-interactions of the gravity-inflaton system. Section 5 gives a two 
loop effect from the potential of a massless, minimally coupled scalar. 
Our conclusions comprise the final section.

\section{Gauge-Fixed Lagrangian}

The model we consider consists of three fields: the spacelike, $D$-dimensional 
metric $\mathbf{g}_{\mu\nu}$; the scalar inflaton $\varphi$ whose slow roll 
down its potential $V(\varphi)$ drives inflation; and a spectator scalar 
$\sigma$ which is centered at the $\sigma_0 = 0$ minimum (of zero) of its 
massless potential $U(\sigma)$. The Lagrangian is,
\begin{eqnarray}
\lefteqn{\mathcal{L} = \Bigl[ \frac{\mathbf{R}}{16 \pi G} - \frac12 
\varphi_{,\mu} \varphi_{,\nu} \mathbf{ g}^{\mu\nu} - V(\varphi) } \nonumber \\
& & \hspace{3cm} - \frac12 \sigma_{,\mu} \sigma_{,\nu} \mathbf{g}^{\mu\nu} - 
U(\sigma) \Bigr] \sqrt{-\mathbf{g}} \; , \qquad
\end{eqnarray}
where $\mathbf{R}$ is the $D$-dimensional Ricci scalar and a comma denotes 
ordinary differentiation. 

We decompose $\mathbf{g}_{\mu\nu}$ into lapse, shift and spatial metric
according to Arnowitt, Deser and Misner (ADM) \cite{ADM},
\begin{equation}
\mathbf{g}_{\mu\nu} dx^{\mu} dx^{\nu} = -N^2 dt^2 + g_{ij} (dx^i \!-\!
N^i dt) (dx^j \!-\! N^j dt) \; .
\end{equation}
ADM long ago showed that the Lagrangian has a very simple dependence upon 
the lapse \cite{ADM},
\begin{equation}
\mathcal{L} = \Bigl({\rm Surface\ Terms}\Bigr) - \frac{\sqrt{g}}{16 \pi G}
\Bigl[N \cdot A + \frac{B}{N}\Bigr] \; . \label{simple}
\end{equation}
The quantity $A$ is a potential energy,
\begin{equation}
A = - R + 16 \pi G \Bigl[V(\varphi) + U(\sigma) + \frac12 g^{ij}
\Bigl(\varphi_{,i} \varphi_{,j} + \sigma_{,i} \sigma_{,j}\Bigr)\Bigr] , 
\label{A}
\end{equation}
where $R$ is the $(D-1)$-dimensional Ricci scalar formed from $g_{ij}$.
The quantity $B$ in (\ref{simple}) is a sort of kinetic energy,
\begin{equation}
B = (E^i_{~i})^2 - E^{ij} E_{ij} - 8\pi G \Bigl[ \Bigl(\dot{\varphi}
-\varphi_{,i} N^i\Bigr)^2 \!\!+\! \Bigl(\dot{\sigma} - \sigma_{,i} N^i\Bigr)^2
\Bigr] , \label{B}
\end{equation}
where $E_{ij}/2N$ is the extrinsic curvature,
\begin{equation}
E_{ij} \equiv \frac12 \Bigl[ N_{i ; j} + N_{j ; i} - \dot{g}_{ij}\Bigr] \; ,
\end{equation}
and a semi-colon denotes covariant differentiation. Varying (\ref{simple})
with respect to $N$ produces an algebraic equation,
\begin{equation}
A - \frac{B}{N^2} = 0 \qquad \Longrightarrow \qquad 
N = \sqrt{ \frac{B}{A}} \;
\end{equation}
This gives the constrained Lagrangian a ``virial'' form,
\begin{equation}
\mathcal{L}_{\rm const} = \Bigl({\rm Surface\ Terms}\Bigr) -
\frac{\sqrt{g}}{8 \pi G} \, \sqrt{A B} \; . \label{virial}
\end{equation}

Further progress requires the use of perturbation theory. The nonzero
background fields are $g_{ij} = a^2(t) \delta_{ij}$ and $\varphi =
\varphi_0(t)$. The two nontrivial Einstein equations can be used to
eliminate the background scalar,
\begin{equation}
\dot{\varphi}_0^2 = -\frac{(D \!-\!2)}{8 \pi G} \, \dot{H} \; , \;
V(\varphi_0) = \frac{(D \!-\!2)}{16 \pi G} \Bigl[ \dot{H} + (D\!-\!1)
H^2\Bigr] .
\end{equation}
Note that the background values of the potential and kinetic terms are equal,
$A_0 = B_0 = (D \!-\!2) [ \dot{H} + (D \!-\!1) H^2]$. Hence the background 
value of the lapse is unity.

We fix time as Maldacena \cite{JM} and Weinberg \cite{SW1},
\begin{equation}
G_0(t,\vec{x}) \equiv \varphi(t,\vec{x}) - \varphi_0(t) = 0 \; . \label{G0}
\end{equation}
The other $(D-1)$ conditions have to do with how we define 
the unimodular part of the metric $\widetilde{g}_{ij}$,
\begin{equation}
g_{ij} = a^2(t) e^{2 \zeta(t,\vec{x})} \widetilde{g}_{ij}(t,\vec{x}) 
\; \Longrightarrow \; \sqrt{g} = a^{D-1} e^{(D-1) \zeta} \; . \label{conf}
\end{equation}
We require $\widetilde{g}_{ij} \equiv \delta_{ij} + h_{ij}$ to be transverse,
\begin{equation}
G_i(t,\vec{x}) \equiv \partial_j \widetilde{g}_{ij}(t,\vec{x}) = 
\partial_j h_{ij}(t,\vec{x}) = 0 \; . \label{Gi}
\end{equation}
(Maldacena and Weinberg imposed transversality on the logarithm of 
$\widetilde{g}_{ij}$.) The resulting Faddeev-Popov determinant depends 
only on $h_{ij}$, and is singular for $\epsilon = 0$.

Of course no gauge can eliminate physical inflatons; with condition 
(\ref{G0}) that degree of freedom resides in $\zeta(t,\vec{x})$. 
Linearized gravitons are carried by $h_{ij}(t,\vec{x})$, and spectator 
scalars are in $\sigma(t,\vec{x})$. By contrast, the shift field 
$N^i(t,\vec{x})$ is a constrained variable which mediates interactions 
between the other fields.

To reach a perturbative form we first employ (\ref{conf}) to exhibit
how the potential (\ref{A}) depends on $\zeta$, $h_{ij}$ and $\sigma$,
\begin{equation}
A = A_0 - R + 16 \pi G \Bigl[ U(\sigma) + \frac{e^{-2 \zeta}}{2 a^2}
\widetilde{g}^{ij} \sigma_{,i} \sigma_{,j}\Bigr] \equiv A_0 (1 + \alpha) 
\; . \label{Aexp}
\end{equation}
Here the spatial Ricci scalar is,
\begin{equation}
R = \frac{e^{-2\zeta}}{a^2} \Bigl[ \widetilde{R} - 2 (D \!-\!2) 
\widetilde{\nabla}^2 \zeta - (D\!-\!2) (D\!-\!3) \zeta^{,k} \zeta_{,k}
\Bigr] \; ,
\end{equation}
where $\widetilde{R} = O(h^2)$ is the Ricci scalar formed from 
$\widetilde{g}_{ij}$ and $\widetilde{\nabla}^2 \equiv \partial_i 
\widetilde{g}^{ij} \partial_j$ is the covariant scalar Laplacian. 
At this stage we can also recognize that $\mathcal{R}$ is just $\zeta$, 
in $D=4$ dimensions and to linearized order \cite{LL},
\begin{equation}
\mathcal{R}(t,\vec{x}) \equiv -\frac{a^2(t)}{4 \nabla^2} \, R 
= \Bigl(\frac{D\!-\!2}{2}\Bigr) \zeta(t,\vec{x}) + O\Bigl(\zeta^2,\zeta h, 
h^2\Bigr) \; . \label{Rdef}
\end{equation}

The kinetic energy (\ref{B}) can be expressed as,
\begin{eqnarray}
\lefteqn{B = A_0 + 2 (D\!-\!2) H \Bigl[ (D\!-\!1) (\dot{\zeta}
\!-\! \zeta_{,k} \widetilde{N}^k) \!-\! \widetilde{N}^k_{~ ,k}
\Bigr] } \nonumber \\
& & \hspace{.3cm} + (D\!-\!2) (\dot{\zeta} \!-\! \zeta_{,k}
\widetilde{N}^k) \Bigl[ (D\!-\!1) (\dot{\zeta} \!-\! \zeta_{,k} 
\widetilde{N}^k) \!-\! 2 \widetilde{N}^k_{~ ,k}\Bigr] \nonumber \\
& & \hspace{1cm} + (\widetilde{N}^{k}_{~ ,k})^2 - \widetilde{E}^{k\ell}
\widetilde{E}_{k\ell} - 8\pi G ( \dot{\sigma} \!-\! \sigma_{,k}
\widetilde{N}^k)^2 \; , \qquad \\
& & \equiv A_0 (1 \!+\! \beta) \; .
\end{eqnarray}
Here we define $\widetilde{N} \equiv N^i$, $\widetilde{N}_i \equiv
\widetilde{g}_{ij} \widetilde{N}^j$ and $\widetilde{E}_{ij} \equiv 
\frac12 [ \widetilde{N}_{i ; j} + \widetilde{N}_{j ; i} - \dot{h}_{ij} ]$.
The next step is to expand the volume part of the constrained 
Lagrangian in powers of $\alpha$ and $\beta$,
\begin{eqnarray}
\lefteqn{- \frac{\sqrt{g}}{8 \pi G} \, \sqrt{A B} = -\frac{a^{D-1} e^{(D-1) 
\zeta}}{8\pi G} \, A_0 \sqrt{ (1 \!+\! \alpha) (1 + \beta)} } \\
& & \hspace{-.4cm} = -\frac{a^{D-1} e^{(D-1)\zeta}}{8\pi G} \, A_0 
\Bigl\{1 \!+\! \frac{(\alpha \!+\! \beta)}{2} \!-\! \frac{(\alpha \!-\! 
\beta)^2}{8} \!+\! \dots \Bigr\} . \qquad \label{expansion}
\end{eqnarray}
As Weinberg noted, the terms involving no derivatives of the gravity 
fields sum up to a total derivative \cite{SW1}. Another important fact is 
that quadratic mixing between $\widetilde{N}^i$ and $\zeta$ can be 
eliminated with the covariant field redefinition,
\begin{equation}
\widetilde{S}^k \equiv \widetilde{N}^k + \widetilde{g}^{k\ell} 
\partial_{\ell} \frac1{\widetilde{\nabla}^2} \Bigl[ \frac{e^{-2\zeta}}{H a^2}
\widetilde{\nabla}^2 \zeta - \epsilon (\dot{\zeta} \!-\! \zeta_{,i}
\widetilde{N}^i)\Bigr] \; .
\end{equation}

After much work the quadratic Lagrangians emerge,
\begin{eqnarray}
\mathcal{L}^{(2)}_S &\! =\! & \frac{a^{D-1}}{32 \pi G} \Bigl\{ \partial_{\ell}
\widetilde{S}^k \partial_{\ell} \widetilde{S}^k \!\!+\! \Bigl( \frac{D \!-\! 
3 \!+\! \epsilon}{D \!-\! 1 \!-\! \epsilon}\Bigr) \partial_{\ell} 
\widetilde{S}^k \partial_k \widetilde{S}^{\ell} \Bigr\} \; , \qquad 
\label{freeN} \\
\mathcal{L}^{(2)}_{\zeta} &\! = \!& \frac{(D \!-\!2) \, \epsilon \, a^{D-1}}{
16\pi G} \Bigl\{ \dot{\zeta}^2 - \frac1{a^2} \partial_k \zeta \partial_k \zeta
\Bigr\} , \label{freeZ} \\
\mathcal{L}^{(2)}_{h} &\! = \!& \frac{a^{D-1}}{64\pi G} \Bigl\{ \dot{h}_{ij}
\dot{h}_{ij} - \frac1{a^2} \partial_k h_{ij} \partial_k h_{ij} \Bigr\} , 
\label{freeh} \\
\mathcal{L}^{(2)}_{\sigma} &\! = \!& \frac{a^{D-1}}{2} \Bigl\{ \dot{\sigma}^2
- \frac1{a^2} \partial_k \sigma \partial_k \sigma\Bigr\} . \label{freephi}
\end{eqnarray}
Expression (\ref{freephi}) reveals $\sigma$ to be a massless, minimally
coupled scalar with unit normalization. Let us call its propagator 
$i\Delta(x;x')$. From (\ref{freeh}), and relations (\ref{conf}-\ref{Gi}), 
we see that the graviton propagator is proportional,
\begin{equation}
i\Bigl[\mbox{}_{ij} \Delta_{k\ell}\Bigr](x;x') = 32\pi G \Bigl[ 
\Pi_{i(k} \Pi_{\ell) j} \!-\! \frac{\Pi_{ij} \Pi_{k\ell}}{D \!-\!2} \Bigr] 
i\Delta(x;x') \; , \label{hprop}
\end{equation}
where $\Pi_{ij} \equiv \delta_{ij} - \partial_i \partial_j/\nabla^2$ is the
transverse projection operator. These relations are exact. Because 
$\epsilon(t) = -\dot{H}/H^2$ is nearly constant during inflation, expression
(\ref{freeZ}) implies a similar relation for the $\zeta$ propagator,
\begin{equation}
i\Delta_{\zeta}(x;x') \approx \frac{8\pi G}{(D \!-\!2) \epsilon} \, 
i\Delta(x;x') \; . \label{zprop}
\end{equation}

The massless, minimally coupled scalar has a well-known infrared problem
\cite{FPVS} which we regulate by working on $T^{D-1}$ with radius $L$
and then making the integral approximation for the mode sum \cite{TWJMPW},
\begin{eqnarray}
\lefteqn{i\Delta(x;x') = \int \!\! \frac{d^{D-1}k}{(2\pi)^{D-1}} \, 
\theta(k \!-\! L^{-1}) e^{i \vec{k} \cdot (\vec{x} - \vec{x}')} } \nonumber \\
& & \hspace{-.3cm} \times \Bigl\{ \theta(t \!-\! t') u(t,k) u^*(t',k) +
\theta(t' \!-\! t) u^*(t,k) u(t',k) \Bigr\} . \qquad \label{modesum}
\end{eqnarray}
The mode function for constant $\epsilon$ is,
\begin{equation}
u(t,k) = \frac{\sqrt{\frac{\pi}{4 (1-\epsilon) H}}}{a^{\frac{D-1}2}} \,
H^{(1)}_{\nu}\Bigl( \frac{k}{(1\!-\!\epsilon) H a}\Bigr) \; , \;
\nu \equiv \frac{D \!-\! 1 \!-\! \epsilon}{2 (1 \!-\! \epsilon)} \; .
\end{equation}
Constant $\epsilon$ implies $H a^{\epsilon}$ is also constant and hence,
\begin{equation}
D-4 = 0 = \dot{\epsilon} \quad \Longrightarrow \quad 
\lim_{t \rightarrow \infty} u(t,k) = C(\epsilon) \times 
\frac{H(t_k)}{\sqrt{2 k^3}} \; , \label{keyeqn}
\end{equation}
where $C(0) = 1$ and we will use $C(\epsilon) \approx 1$ generally.

Relations (\ref{Rdef}), (\ref{zprop}) and (\ref{keyeqn}) allow a trivial
derivation of the typical result (\ref{typR}) for the scalar power spectrum,
\begin{equation}
\Bigl[ \Delta^2_{\mathcal{R}}(k,t) \Bigr]_{\rm tree} \approx
\frac{k^3}{2 \pi^2} \times \frac{8\pi G}{2 \epsilon} \times 
\vert u(t,k)\vert^2 \approx \frac{G H^2(t_k)}{\pi \epsilon} \; . \label{triv}
\end{equation}
The tensor power spectrum is,
\begin{equation}
\Delta^2_{h}(k,t) \equiv \frac{k^3}{2 \pi^2} \int \!\! d^3x \,
e^{-i \vec{k} \cdot \vec{x}} \Bigl\langle \Omega \Bigl\vert h_{ij}(t,\vec{x})
h_{ij}(t,\vec{0}) \Bigr\vert \Omega \Bigr\rangle \; .
\end{equation}
Relations (\ref{hprop}) and (\ref{keyeqn}) offer a similarly
straightforward derivation of the typical tree result for $\Delta_h^2(k,t)$
\cite{LL,reviews},
\begin{equation}
\Bigl[ \Delta^2_{h}(k,t)\Bigr]_{\rm tree} 
\!\!\!\!\!\! = \frac{k^3}{2 \pi^2} \times 32 \pi G \times 
2 \times \vert u(t,k) \vert^2 \!\approx\!  \frac{16}{\pi} 
G H^2(t_k) \; . \label{typh}
\end{equation}
The $1/\epsilon$ enhancement of the scalar power spectrum with regard to
the tensor one presumably explains why the scalar contribution has been
detected but the tensor signal has so far not been resolved. At 95\% 
confidence the bound on their ratio at $k_0 = 0.002~{\rm Mpc}^{-1}$ 
is \cite{WMAP},
\begin{equation}
r \equiv \frac{\Delta^2_{h}(k_0)}{\Delta^2_{\mathcal{R}}(k_0)} < 0.22 
\, . \label{rbound}
\end{equation}
With the typical tree order results (\ref{typR}) and (\ref{typh}), and the
measured spectrum (\ref{data}), this bound implies an upper limit on the 
inflationary Hubble parameter,
\begin{equation}
G H^2(t_{k_0}) \approx \frac{\pi}{16} \times r \times 
\Delta^2_{\mathcal{R}}(k_0) \ltwid 10^{-10} \, .
\end{equation}
We can also get a bound on $\epsilon$ by combining the typical tree results
(\ref{typR}) and (\ref{typh}) with (\ref{rbound}),
\begin{equation}
\epsilon(t_{k_0}) \approx \frac{r}{16} \ltwid 0.014 \; .
\end{equation}
This is why it was justified to use $C(\epsilon) \approx 1$. {\it We shall
go further and approximate loop corrections by taking the de Sitter
limit of $\epsilon = 0$ once multiplicative factors of $\epsilon$
have been removed from vertices and propagators.}

It remains to derive the relevant interactions by expanding the
constrained Lagrangian (\ref{expansion}). These interactions are 
quite complicated, but most of them are precluded by too many 
differentiated fields from contributing the maximum number of 
infrared logarithms. If we want just the maximum possible number
of infrared logarithms then the number of interactions at any
order becomes manageable. To study the lowest order effects of $\zeta$ 
self-interactions it suffices to consider the minimal generalization
of (\ref{freeZ}),
\begin{equation}
\mathcal{L}_{\zeta} = \frac{(D \!-\!2) \, \epsilon}{16 \pi G} \, a^{D-1}
e^{(D-1) \zeta} \Bigl\{ \dot{\zeta}^2 - \frac{e^{-2\zeta}}{a^2} 
\partial_k \zeta \partial_k \zeta \Bigr\} \; . \label{zint}
\end{equation}
To study the lowest order effects of the scalar potential we need only,
\begin{equation}
\mathcal{L}_{U} = \frac{\epsilon}{D \!-\! 1} \, a^{D-1} e^{(D-1) \zeta}
U(\sigma) \; . \label{Uint}
\end{equation}

\section{Infrared Logarithms}

Infrared logarithms are factors of $\ln[a(t)]$ which can contaminate
loop corrections involving undifferentiated gravitons or massless, 
minimally coupled scalars. The oldest example is from 1982 \cite{VFLS} 
and consists of the coincidence limit of $i\Delta(x;x')$ on de Sitter 
background ($\epsilon(t) = 0$ and $H(t) = H_I$). With full dimensional
regularization, and $L = (H_I a_I)^{-1}$, the result is \cite{TWJMPW},
\begin{eqnarray}
\lefteqn{i\Delta(x;x) = \int \frac{d^{D-1}k}{(2\pi)^{D-1}} \, 
\theta(k \!-\! H_I a_I) \vert u(t,k) \vert^2 \; , } \\
& & = \frac{a^{-(D-1)}}{2^D \pi^{\frac{D-3}2} \Gamma(\frac{D-1}2)}
\int_{H_I a_I}^{\infty} \!\!\!\!\!\!\! dk \, k^{D-2} \Bigl\vert 
H^{(1)}_{\frac{D-1}2}\Bigl( \frac{k}{H_I a}\Bigr) \Bigr\vert^2 \; , 
\qquad \label{line2} \\
& & = \frac{H_I^{D-2}}{2^D \pi^{\frac{D-3}2} \Gamma(\frac{D-1}2)}
\int_{a_I a^{-1}}^{\infty} \!\!\!\!\!\!\! dz \, z^{D-2} \Bigl\vert 
H^{(1)}_{\frac{D-1}2}( z) \Bigr\vert^2 \; , \label{line3} \\
& & = \frac{H_I^{D-2}}{ (4 \pi)^{\frac{D}2}} \frac{\Gamma(D \!-\!1)}{
\Gamma(\frac{D}2)} \Bigl\{ 2 \ln\Bigl[ \frac{a(t)}{a_I}\Bigr]
-\psi\Bigl(1 \!-\! \frac{D}2\Bigr) \nonumber \\
& & \hspace{.8cm} + \psi\Bigl( \frac{D \!-\!1}2\Bigr) + \psi(D \!-\!1)
+ \psi(1) + O\Bigl(\frac{a_I^2}{a^2}\Bigr) \Bigr\} . 
\qquad \label{fcoin}
\end{eqnarray}
This exhibits the fallacy of the argument Senatore and Zaldarriaga 
gave against infrared logarithms based on ``making the integral 
dimensionless'' \cite{LSMZ}. That is the change of variables from $k$ 
to $z = k/H_I a(t)$ in passing from (\ref{line2}) to (\ref{line3}). Had
the lower limit been $k = 0$ this would indeed have eliminated 
any time dependence, however, the integral would have been infrared
divergent. So we come to a crucial insight: {\it infrared logarithms derive
from diagrams that would be infrared divergent as in-out matrix elements 
on the spatial manifold $R^{D-1}$.} Of course that is why they are called
infrared logarithms.

Note that any derivatives, with respect to space or time, would have 
eliminated the infrared logarithm in (\ref{fcoin}). This observation has 
led to a very simple rule for inferring the maximum number of infrared 
logarithms which can come from a particular interaction \cite{TW1}: 
{\it If the interaction has a total of $K$ undifferentiated gravitons
and undifferentiated massless, minimally coupled scalars, after partial
integration has been exhausted, then each correction involving two such 
interactions can produce as many as $K$ infrared logarithms.}

This rule has been tested in a variety of explicit, fully dimensionally
regulated and renormalized computations on de Sitter background using
the Schwinger-Keldysh formalism \cite{JSKTMBMLVK}. For a massless, minimally 
coupled scalar with a quartic self-interaction the rule correctly predicts 
the number of infrared logarithms in the expectation value of the stress 
tensor at one and two loop orders \cite{OW}, and in the one and two loop 
order self-mass-squared \cite{BOW,KO}. For scalar quantum electrodynamics
the rule correctly predicts the infrared logarithms which are seen in the
one loop vacuum polarization \cite{PTW} and in the two loop scalar and
electrodynamic field strengths, as well as the two loop expectation value
of the stress tensor \cite{PTsW}. For a Yukawa-coupled scalar the rule
has been checked with the one loop fermion self-energy \cite{PW} and with
the expectation value of the coincident vertex at two loop order \cite{MW1}.
The rule also gives the correct number of infrared logarithms in the one
loop fermion self-energy from quantum gravity \cite{MW2}.

Senatore and Zaldarriaga considered two models in detail. The first, given
in their equation (11) and studied in section 3, consists of a Lagrangian
containing only differentiated fields \cite{LSMZ}. Any such interaction
has $K=0$, so the rule predicts no infrared logarithms, which is what they 
found. Their second model, given in their equation (75) and studied in 
section 4, was the same as Weinberg's: gravity + inflaton + $N$ massless, 
minimally coupled scalars  with no potential \cite{LSMZ}. This model shows 
infrared logarithms, both from its $\zeta$ self-interactions and from 
interactions with undifferentiated $h_{ij}$ fields. However, Senatore and 
Zaldarriaga ignored those interactions because they give no parametric 
enhancement involving the potentially large number $N$. The scalar kinetic 
terms which were the object of their study consist of differentiated
$\sigma$ fields with a complicated set of couplings to one $\zeta$ field.
Although there are certainly some $K=1$ terms present, cancellations make
the resulting integrals infrared finite \cite{SW1,LSMZ} so the rule again 
predicts no infrared logarithms, and that is what they found.

\begin{figure}
\includegraphics[width=2.4cm,height=1.8cm]{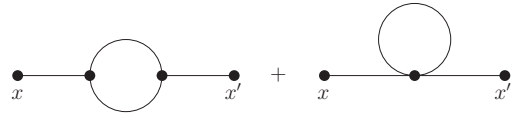}
\caption{One loop correction from cubic and quartic self-interactions
of $\zeta$.}
\label{zfig1}
\end{figure}

\begin{figure}
\begin{center}
\includegraphics[width=2.4cm,height=1.8cm]{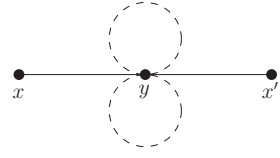}
\end{center}
\caption{Two loop correction from the interaction given in expression
(\ref{Uint}). Solid lines represent $\zeta$ and dashed lines represent
$\sigma$.}
\label{zfig2}
\end{figure}

Let us now consider the two interactions (\ref{zint}) and (\ref{Uint})
given in the previous section. The general form of the $\zeta$
self-interaction (\ref{zint}) is $\zeta^K \partial \zeta \partial \zeta$,
which is the same as for quantum gravity. We therefore expect that there
should be a single infrared logarithm from a correction involving two
3-point interactions, with $K=1$, or from a single 4-point interaction,
with $K=2$. These corrections correspond to the diagrams depicted in 
Fig.~\ref{zfig1}, and we will show in the next section that they indeed
produce a single infrared logarithm. Supposing that the spectator 
potential is quartic, we see that (\ref{Uint}) contains an interaction
of the form $\zeta^2 \sigma^4$. This has $K=6$, so the rule predicts 
three infrared logarithms from a correction which involves one such
interaction. The corresponding diagram is depicted in Fig.~\ref{zfig2} 
and we will confirm that it does produce three infrared logarithms in
the penultimate section.

\section{Time Dep. from Self-Interactions of $\zeta$}

The two diagrams of Fig.~\ref{zfig1} derive from expanding expression
(\ref{zint}) to cubic and quartic orders. The second of these diagrams
is very similar to the computation featured in section V of Weinberg's
paper \cite{SW1}. As he noted, a field redefinition would make (\ref{zint}) 
free were it not for the extra factor of $e^{-2\zeta}$ on the term with
space derivatives. Things can be simplified by exploiting Weinberg's 
observation that only the time derivative term contributes an infrared 
logarithm at one loop order \cite{SW1}. We therefore make the field
redefinition,
\begin{equation}
Z \equiv \frac2{D\!-\!1} \Bigl[ e^{\frac{D-1}2 \zeta} \!-\! 1\Bigr] \;
\Leftrightarrow \; \zeta = \frac2{D \!-\!1} \ln\Bigl[1 \!+\! \frac{D-1}2 \, 
Z\Bigr] \; , \label{redef}
\end{equation}
and forget about the residual interactions involving spatial derivatives.

The $Z$ propagator is the same as the $\zeta$ propagator (\ref{zprop}).
Hence the one $\zeta$ loop correction to the $\zeta$--$\zeta$ correlator is,
\begin{eqnarray}
\lefteqn{ \Bigl\langle \Omega \Bigl\vert \zeta(x) \zeta(x') \Bigr\vert
\Omega \Bigr\rangle_{\zeta \, {\rm loop}} \approx \Bigl( \frac{D \!-\! 1}2
\Bigr)^2 \Bigl\langle \Omega \Bigl\vert \frac13 Z(x) Z^3(x') } \nonumber \\
& & \hspace{1.8cm} + \frac14 Z^2(x) Z^2(x') + \frac13 Z^3(x) Z(x')
\Bigr\vert \Omega \Bigr\rangle \; , \qquad \\
& & \approx \Bigl(\frac{D \!-\! 1}2\Bigr)^2 \Bigl[ \frac{8\pi G}{(D \!-\! 
2) \epsilon}\Bigr]^2 \Bigl\{ i\Delta(x;x') i\Delta(x';x') \nonumber \\
& & \hspace{1.8cm} + \frac12 [ i\Delta(x;x')]^2 \!+ i\Delta(x;x) 
i\Delta(x;x')\Bigr\} . \qquad \label{props}
\end{eqnarray}
We now set $x^{\mu} = (t,\vec{x})$ and $x^{\prime \mu} = (t,\vec{0})$, and 
Fourier transform on $\vec{x}$. It also makes sense to retain only the 
infrared logarithm terms because time independent contributions derive as 
well from derivative interactions of the same order as (\ref{zint}) which
we have ignored. That is where the ultraviolet divergences reside, and 
they can be absorbed into BPHZ counterterms as usual. In the absence of
any condition for fixing the finite parts of those counterterms, the
infrared logarithm terms are the only unambiguous prediction. The final 
result is,
\begin{equation}
\Bigl[\Delta^2_{\mathcal{R}}(k,t) \Bigr]_{\zeta\, {\rm loops}} \!\!\!\!\!
\approx \frac{G H^2}{\pi \epsilon} \Bigl\{ \frac{27 G H^2}{4 \pi \epsilon}
\, \ln(a) + O(G^2 H^4) \Bigr\} . \label{selfcor}
\end{equation}

We should mention that it is by no means clear what collection of fields 
represents the observed scalar power spectrum $\Delta^2_{\mathcal{R}}(k,t)$. 
At tree order it suffices to use the $\zeta$--$\zeta$ correlator, and our 
result (\ref{selfcor}) is based on extending that correspondence to all 
orders. This definition affords a simple renormalization scheme because then
the power spectrum is a noncoincident Green's function of a fundamental 
field, and ordinary renormalization makes those finite. However, it is 
conceivable that the measured quantity is actually the correlator of some 
composite operator such as (\ref{Rdef}), in which case an additional, 
composite operator renormalization would be required. Nonlinear modifications 
of the observable can introduce additional infrared logarithms. For example, 
if the correct observable is the correlator of the field $Z(t,\vec{x})$ 
defined in expression (\ref{redef}), then there are no infrared logarithms 
at one loop order. However, there does not seem any reason to suppose this, 
and even doing so would not prevent the appearance of infrared logarithms at 
higher orders.

\section{Time Dep. from Spectator Potentials}

Let us assume $U(\sigma) = \lambda \sigma^4/4!$. The diagram of 
Fig.~\ref{zfig2} derives from the $\zeta^2 \sigma^4$ term of the
interaction (\ref{Uint}),
\begin{equation}
\Delta \mathcal{L} = \frac{(D-1)}{48} \, \lambda \epsilon a^{D-1}
\zeta^2 \sigma^4 \; . \label{DL}
\end{equation}
The Schwinger-Keldysh \cite{JSKTMBMLVK} result for this diagram is,
\begin{eqnarray}
\lefteqn{({\rm Fig.~\ref{zfig2}}) \approx \Bigl[ \frac{8\pi G}{(D \!-\!2)
\epsilon}\Bigr]^2 \int \! d^Dy \Bigl\{ i\Delta_{\scriptscriptstyle ++}(x;y)
i\Delta_{\scriptscriptstyle ++}(x';y) } \nonumber \\
& & \hspace{-.5cm} - i\Delta_{\scriptscriptstyle +-}(x;y) 
i\Delta_{\scriptscriptstyle +-}(x';y) \Bigr\} \frac{i\lambda (D\!-\!1)}{8}
\, \epsilon a^{D-1} [ i\Delta(y;y)]^2 \! . \qquad \label{stp1}
\end{eqnarray}
The propagator $i\Delta_{\scriptscriptstyle ++}(x;x')$ is the same mode
sum as (\ref{modesum}), whereas $i\Delta_{\scriptscriptstyle +-}(x;x')$ has 
the same first line as (\ref{modesum}) but the curly-bracketed expression 
on the second line is replaced by just $u^*(t,k) u(t',k)$.

We again take $x^{\mu} = (t,\vec{x})$ and $x^{\prime \mu} = (t,\vec{0})$,
and Fourier transform on $\vec{x}$, to obtain,
\begin{eqnarray}
\lefteqn{ \int \! d^{D-1}\!x \, e^{-i \vec{k} \cdot \vec{x}}
({\rm Fig.~\ref{zfig2}}) = \frac{i 8 \pi^2 (D\!-\!1) \lambda G^2}{(D-2)^2
\epsilon} \! \int_0^t \!\! ds \, [a(s)]^{D-1} } \nonumber \\
& & \times \Bigl\{ [u(t,k)]^2 [u^*(s,k)]^2 \!-\! 
[u^*(t,k)]^2 [u(s,k)]^2 \Bigr\} [i\Delta]^2 . \qquad \label{stp2}
\end{eqnarray}
There is no point in retaining the divergent part of the coincident 
propagator (\ref{fcoin}), and continuing to work in $D$ dimensions,
unless we add the various counterterm diagrams. That exercise is identical
to the published two loop computation of the expectation value of the
$\sigma$ stress tensor \cite{OW}. We will therefore retain only the 
leading infrared logarithm terms and take $D=4$. 

Oscillations of the mode functions preclude a coherent effect before 
first horizon crossing. After horizon crossing one may take the long 
wavelength limit of the mode functions,
\begin{equation}
u(t,k) \longrightarrow \frac{H}{\sqrt{2 k^3}} \Bigl\{ 1 + \frac12
\Bigl(\frac{k}{H a}\Bigr)^2 + \frac{i}3 \Bigl(\frac{k}{H a}\Bigr)^3
+ \dots\Bigr\} .
\end{equation}
Hence the curly-bracketed term of (\ref{stp2}) becomes,
\begin{eqnarray}
\lefteqn{\Bigl\{ [u(t,k)]^2 [u^*(s,k)]^2 \!-\! [u^*(t,k)]^2 
[u(s,k)]^2 \Bigr\} } \nonumber \\
& & \hspace{1cm} \longrightarrow -\frac{iH}{k^3} \Bigl\{ \frac13
\Bigl[ \frac1{a^3(s)} - \frac1{a^3(t)} \Bigr] + O\Bigl(\frac{k^2}{H^2}
\Bigr) \Bigr\} . \qquad
\end{eqnarray}
Putting everything together produces,
\begin{eqnarray}
\lefteqn{\Bigl[\int \! d^{D-1}\!x \, e^{-i \vec{k} \cdot \vec{x}} 
({\rm Fig.~\ref{zfig2}}) \Bigr]_{\rm leading\ log} } \nonumber \\
& & \hspace{1.5cm} \approx \frac{\lambda G^2 H^5}{8 \pi^2 \epsilon k^3} 
\int_{t_k}^t \!\! ds \, \Bigl[1 - \frac{a^3(s)}{a^3(t)}\Bigr] \ln^2[a(s)]
\; , \qquad \\
& & \hspace{1.5cm} = \frac{\lambda G^2 H^4}{24 \pi^2 \epsilon k^3} 
\Bigl\{ \ln^3[a(t)] + {\rm subleading} \Bigr\} \; . \qquad
\end{eqnarray}
And multiplying by $k^3/2\pi^2$ gives the power spectrum,
\begin{equation}
\Bigl[\Delta^2_{\mathcal{R}}(k,t) \Bigr]_{\sigma\, {\rm loops}} \!\!\!\!\!
\approx \frac{G H^2}{\pi \epsilon} \Bigl\{ \frac{\lambda G H^2}{48 \pi^3}
\, \ln^3(a) + O(\lambda^2) \Bigr\} . \label{Ucor}
\end{equation}

\section{Conclusions}

We have shown that the $\zeta$--$\zeta$ correlator acquires time
dependent infrared log corrections, starting at one loop (\ref{selfcor})
from $\zeta$ self-interactions, and at two loops (\ref{Ucor}) from the 
quartic potential of a spectator scalar. There should also be infrared 
logarithms from dynamical gravitons, starting at two loops. 

The physical interpretation of the spectator effect (\ref{Ucor}) derives
from the small increase in the vacuum energy as inflationary particle 
production pushes the $\sigma$ field up its potential $U(\sigma)$. The
dimensionally regulated and fully renormalized result for this has been
derived \cite{OW}, but we can understand its effect on the $\zeta$ mode
functions by simply adding the Hartree approximation of (\ref{DL}) to
the free $\zeta$ Lagrangian (\ref{freeZ}) in $D=4$ dimensions,
\begin{equation}
\mathcal{L}^{(2)}_{\zeta} \longrightarrow \frac{\epsilon a^3}{8 \pi G} 
\Bigl\{ \dot{\zeta}^2 - \frac1{a^2} \partial_k \zeta \partial_k \zeta + 
\frac{3 \lambda G H^4}{32 \pi^3} \, \ln^2(a) \zeta^2 \Bigr\} \; .
\end{equation}
After horizon crossing the associated mode equation is,
\begin{eqnarray}
\lefteqn{3 H \dot{u} \approx \frac{3 \lambda G H^4}{32 \pi^3} \, \ln^2(a) 
u \; } \nonumber \\
& & \hspace{1cm} \Longrightarrow u(t,k) \approx \frac{H}{\sqrt{2 k^3}} 
\Bigl\{1 + \frac{\lambda G H^2}{96 \pi^3} \, \ln^3(a) \Bigr\} \; . \qquad
\end{eqnarray}
Inserting the quantum corrected mode function in expression (\ref{triv})
gives precisely our result (\ref{Ucor}). It seems likely that a similar
explanation can be given for the effects from $\zeta$ self-interactions, 
and from interactions with gravitons.

Such effects must be present or else there is something seriously wrong 
with our understanding of how gravity responds to quantum fluctuations.
That they would even be questioned is a tribute to how firmly cosmologists
have come to believe in the time independence of $\widetilde{\zeta}(t,
\vec{k})$ after horizon crossing. Of course we appreciate the wonder of 
preserving a memory of conditions from inflation, but the practical value 
of $\Delta^2_{\mathcal{R}}(k,t)$ does not seem compromised by the 
minuscule time dependence we have exhibited. The loop corrections we have 
discussed can never be large (which is the same conclusion reached by
Weinberg \cite{SW2,Bua}) because they are suppressed by the quantum 
gravitational loop counting parameter $G H^2 \ltwid 10^{-10}$. Their 
enhancement by $\ln[a(t)/a(t_k)] \ltwid 60$ is huge by the standards of 
conventional perturbation theory, and unprecedented in view of its time 
dependence, but there are simply not enough e-foldings of inflation left 
after first horizon crossing to overcome the suppression factor for any 
mode whose spatial variation we can now perceive.

Despite having reached a different conclusion from Senatore and Zaldarriaga,
our results represent no real disagreement with their analysis. They were 
uninterested in self-interactions from the gravity-inflaton system because 
the fixed number of fields in that sector cannot engender effects which are 
enhanced by a potentially large parameter such as Weinberg's $\mathcal{N}$.
And they dismissed massless scalars with nonzero potentials as unnatural. 
We feel it is not reasonable to fine tune the inflaton potential $V(\varphi)$ 
and then quibble about fine tuning the spectator potential $U(\sigma)$. We 
also thought it worth establishing that infrared logarithms do contaminate 
gauge invariant quantum gravity observables such as $\Delta^2_{
\mathcal{R}}(k,t)$ because the contrary view has been expressed 
\cite{GarTan,TWN}.

We close with two thoughts. First, the small infrared log corrections to 
$\Delta^2_{\mathcal{R}}(k,t)$ might eventually be observable through 
21 centimeter measurements of the matter power spectrum out to very large 
redshifts \cite{21cm}. This would require untangling the primordial signal 
from late time effects, which is very hard but perhaps not impossible. It 
would also require a precise tree order prediction from some unique model 
of inflation.

Our final comment is that loop corrections to the power spectrum are not 
the best place to study infrared logarithms because $\ln[a(t)/a(t_k)]$
cannot exceed about 60 for any mode whose spatial variation we now perceive.
By contrast, there can be spectacular enhancements in quantities which seem
spatially constant, such as the vacuum energy \cite{TWa} and Newton's 
constant \cite{TWb}, because they receive contributions from modes which are 
still super-horizon. For a very long period of inflation perturbation theory 
can even break down, after which reliable computations would require some 
nonperturbative resummation technique. Such a method has been devised by 
Starobinsky \cite{AAS}, and applied by him and Yokoyama to scalar potential 
models \cite{SY}, for which it sums the series of leading infrared 
logarithms \cite{TW1}. Starobinsky's method has recently been extended to 
Yukawa-coupled fermions \cite{MW1} and to scalar quantum electrodynamics 
\cite{PTsW3}. It has not yet been extended to quantum gravity but there are 
reasons for believing that some version of it can be \cite{MW3}.

\begin{acknowledgments}

We are grateful for correspondence with S. Deser and S. Weinberg. This 
work was partially supported by Marie Curie Grant IRG-247803, by NSF 
grants PHY-0653085 and PHY-0855021, and by the Institute for Fundamental 
Theory at the University of Florida.

\end{acknowledgments}

\end{document}